# null²: Boundary-Dissolving Bodies and Architecture towards Digital Nature

Yoichi Ochiai

Research and Development Center for Digital Nature, University of Tsukuba, wizard@slis.tsukuba.ac.jp

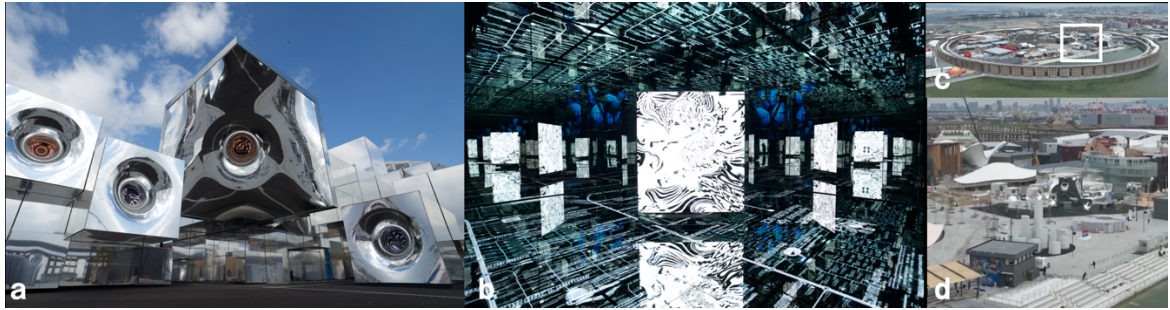

Figure 1: (a) Exterior view, (b) Interior view, (c) Pavilion location within the overall site, (d) Enlarged view of (c).

This paper presents a case study of the thematic pavilion null² at Expo 2025 Osaka-Kansai, in contrast to the symbolic structure based on static Jomon motifs epitomized by Taro Okamoto's Tower of the Sun from Expo '70. The study discusses the Yayoi-inspired mirror motifs and dynamically transforming interactive spatial configuration of null², where visitors themselves become integrated as experiential content. The shift from traditional static representation to a new ontological and aesthetic model, characterized by the visitor's body merging in real-time with architectural space at an installation scale, is analyzed. Referencing the philosophical context of the Expo '70 theme "Progress and Harmony for Mankind," this research reconsiders the worldview articulated by null² in Expo 2025, in which computation is naturalized and ubiquitous, through its intersection with Eastern philosophical traditions. Specifically, it investigates how the immersive experiences within the pavilion—grounded in the philosophical framework of Digital Nature—reinterpret traditional spatial and structural motifs of the tea room, positioning them within the contemporary discourse of digital art installations. The aim of this study is to contextualize and document null² as an important contemporary case study arising from Expo practices, considering the historical and social background in Japan from the 19th to the 21st century, during which world expositions have served as pivotal points for the birth of the modern Japanese concept of "fine art," symbolic milestones of economic development, and key moments in urban and media culture formation. Furthermore, this paper academically organizes the architectural techniques, computer graphics methodologies, media art practices, and the theoretical backgrounds utilized in null², highlighting the scholarly significance of preserving these as an archival document for future generations.

CCS Concepts: · Applied computing;Arts and humanities; Media arts;

**Additional Key Words and Phrases:** Digital Nature, Computational Architecture, Media Art, Expo 2025, Human-Computer Interaction,  Mirrored Body, Blockchain, AI, Eastern Philosophy, Object-Oriented Programming

## 1 INTRODUCTION

For Japan, international expositions (World Expos) have historically served as pivotal moments marking the modern establishment of the concept of "art" within the country. Japan's first participation in the 1867 Paris Exposition introduced traditional Japanese crafts and ukiyo-e woodblock prints to European art circles, igniting the international cultural phenomenon known as "Japonisme" [1]. Additionally, the 1970 Japan World Exposition (Expo '70) held in Osaka was symbolized by Taro Okamoto's "Tower of the Sun," which critically engaged with technological optimism and progressivism prevalent during Japan's rapid economic growth, thereby redefining Japanese cultural and spiritual identity. Okamoto's primitive yet futuristic expression occupies a significant position in art and architectural history due to its critical stance between national narratives and internationalist ideals [2].

Against this historical backdrop, the World Expo returns to Osaka-Kansai in 2025. The thematic pavilion project addressed in this paper, "null²," is a response to the static Jōmon-inspired motifs of the 1970 Expo's "Tower of the Sun," employing Yayoi-period mirror symbolism and dynamically transforming interactive architectural technologies [3] shown in figure 1 (a) – (d). Specifically, "null²" transcends traditional static representations by integrating visitors into real-time spatial experiences, proposing a novel aesthetic model as an installation at the architectural scale. Comparisons are shown in figure 2 (a)-(b).

Underlying this new symbolic structure presented by null² is the philosophical concept of "Digital Nature," defined as the fusion of computing technology seamlessly integrated and omnipresent within the natural environment, proposing a new vision of nature beyond anthropocentrism [4][5]. While Expo '70 embodied the utopian ideal of "Progress and Harmony for Mankind" through technological innovation, null² at Expo 2025 responds to the current Expo's theme "Designing Future Society for Our Lives," aiming to merge computation into nature itself and redefine everyday environments and bodily sensations through an Eastern philosophical perspective.

Given this context, the objective of this study is to reaffirm the historical and social significance of World Expos as formative events in Japanese cultural identity and as symbolic milestones in economic development since the 19th century. Furthermore, this paper highlights the scholarly significance of positioning large-scale, ephemeral projects such as international expositions—traditionally difficult to archive academically—as integral subjects within the computer graphics and media art communities. As an initial case-study paper intended to promptly disseminate research insights, it provides a detailed documentation and analysis of the null² project, laying the foundation for a comprehensive academic archive that can support future research. Specifically, this study identifies and articulates null²'s methodological contributions at the intersection of architectural practice and media art, thereby enabling a lasting scholarly reassessment of similar ephemeral, large-scale endeavors.

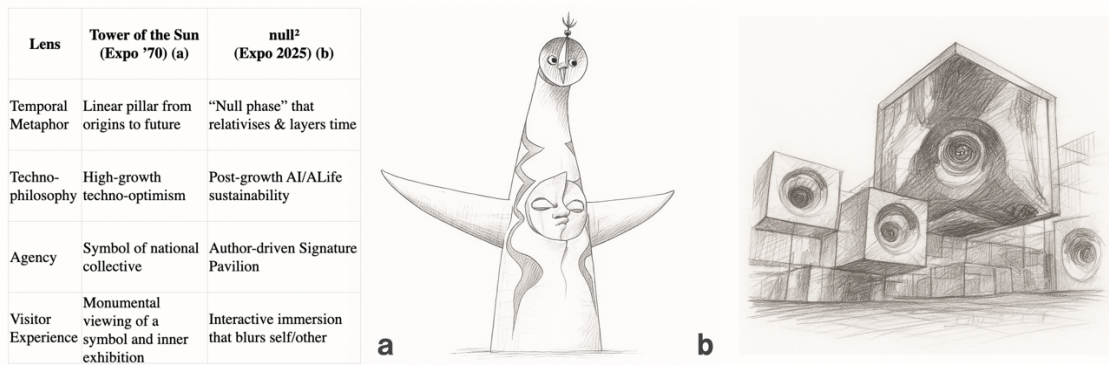

Figure 2: Comparison with the Tower of the Sun from the 1970 Expo, which shared a similar official thematic role. (a) Tower of the Sun (illustration), (b) null² (illustration).

## 2 INSERTING CONTENT ELEMENTS

null² adopts two mirrors as symbolic motifs to address a fundamental architectural challenge commonly faced by World Exposition pavilions: the division between the exterior appearance of the building and its internal exhibition contents. Traditionally, expo pavilions tend to have a fragmented relationship between their external architectural form and internal display, making it difficult to unify symbolic expression and visitor experience [6]. In contrast, null² introduces a double-mirror structure composed of a dynamic exterior mirror membrane and an internal infinite reflection room, conceptually integrating the pavilion's exterior and interior, thus dynamically dissolving boundaries between subject and object, as well as interior and exterior spaces.

In Japanese culture, mirrors have historically held sacred significance, exemplified by the Yata-no-Kagami—one of the "Three Sacred Treasures" central to Shinto rituals [7]. Ancient Japanese revered mirrors as divine mediums, placing them at the heart of shrines and closely associating them with Yayoi-era lifestyles symbolized by rice paddies and agricultural practices [8]. At the 1970 Osaka Expo, Tarō Okamoto's "Tower of the Sun" prominently featured a golden mirror-like face at its center, symbolizing a dialogue between self and other, modernity and primitivism [2]. Building upon such traditional symbolic significance, null² introduces innovative dynamic mirror membranes on its facade, while the internal infinite reflection chamber creates real-time interactions, fusing visitors' bodies with the space and presenting unprecedented expressive potential.

Furthermore, null²'s conceptual foundation is deeply intertwined with Buddhist philosophy, particularly the Heart Sutra's principle of "form is emptiness, emptiness is form" [9]. The pavilion's name, "null²," denotes the layered concept of emptiness repeated twice, merging the Buddhist notion of "emptiness" (空) with the informational concept of "null" in programming languages. This fusion symbolizes the mutual permeation of informational void and Buddhist emptiness. null² visually embodies the fundamental life processes emerging from and returning to nothingness, creating experiences in which visitors themselves become part of this process. Thus, by embedding computational technology seamlessly within natural environments and material worlds, it transcends anthropocentrism to portray an integrated worldview where human and environment, subject and space, mutually interpenetrate.

Within the histories of architecture and media art, numerous works have explored mirror-based expressions. Notably, the Pepsi Pavilion at Expo '70, developed by Experiments in Art and Technology (E.A.T.), incorporated extensive internal mirror surfaces, creating infinite reflections [10]. Similarly, at Expo 2020 Dubai, the Swiss Pavilion featured a fully mirrored facade that visually integrated the cityscape, natural environment, and visitors [11]. However, these examples utilized static mirrors without dynamic real-time deformation.

Various contemporary artworks employing dynamic mirrors also exist. For instance, Mirror Wall [12] presents a single vibrating mirror that subtly distorts each visitor's reflection to trigger perceptual uncertainty, whereas null² scales deformable mirror material into an entire kinetic mirror-membrane façade animated by robotic arms and acoustic waves, embeds LED illumination, and integrates AI-driven avatars—transforming a local perceptual trick into an immersive, data-driven architectural ecosystem. Daniel Rozin's "Wooden Mirror" (1999) employs mechanically actuated wooden pieces to generate abstracted visual reflections, fundamentally different from real-time mirror reflection [13]. null²'s real-time physical and visual mirror deformation thus opens a novel dimension beyond existing works. Positioned as a critical response to historical, philosophical contexts, and related precedents, null² inherits the critical expressions initiated by the "Tower of the Sun" in 1970. By overcoming contemporary technical challenges through dynamic mirror membranes, it introduces a novel and distinctive symbolic representation that elucidates new relationships between computation and nature, humanity, and environment.

## 3 EXPERIENCE

The visitor experience at null² is structured around a seamless integration of digital and physical interactions, beginning even before arriving at the pavilion shown in figure 3 (a)-(i). Visitors initially receive a notification via email, prompting them to download specialized mobile applications, namely Mirrored Body and Scaniverse. These applications allow visitors to capture their personal 3D scan data and generate an interactive AI avatar, incorporating their unique voice and personality characteristics. This digital twin enables interactions with visitors even prior to their actual visit. Moreover, visitors gain deeper comprehension of null²'s conceptual world through additional content provided beforehand, such as illustrated books and comics, facilitating a meaningful dialogue and attachment to their digital self [14, 15].

Upon arrival at the Expo site and approaching the null² pavilion, visitors first encounter the dynamically transforming mirror-membrane façade. Equipped with real-time camera-based sensors, the façade continuously reflects and adapts to the surrounding landscape and visitor movements. Unlike conventional static architectural surfaces, this dynamic façade provides sculptural and interactive visual experiences, bridging the gap between the environment and visitors' movements and cultivating continuously evolving spatial relationships [16].

Inside the pavilion, visitors experience two distinct interaction modes. The first mode, known as Installation Mode, is characterized by a room entirely covered in resonant LED-integrated mirror surfaces (walls, floor, and ceiling). Visitors view this immersive environment from outside the mirrored chamber, observing the visual interplay of artificial life forms through multiple reflections generated by integrated LED arrays placed beneath the mirror surfaces. This approach produces a profoundly immersive perceptual experience, distinguished from established infinity mirror installations, such as Yayoi Kusama's Infinity Mirror Room [17] or teamLab's Borderless [18], by offering an advanced form of bodily and spatial interaction. Visitors experience a narrative of symbolic, boundaryless digital life within this dynamic architectural body.

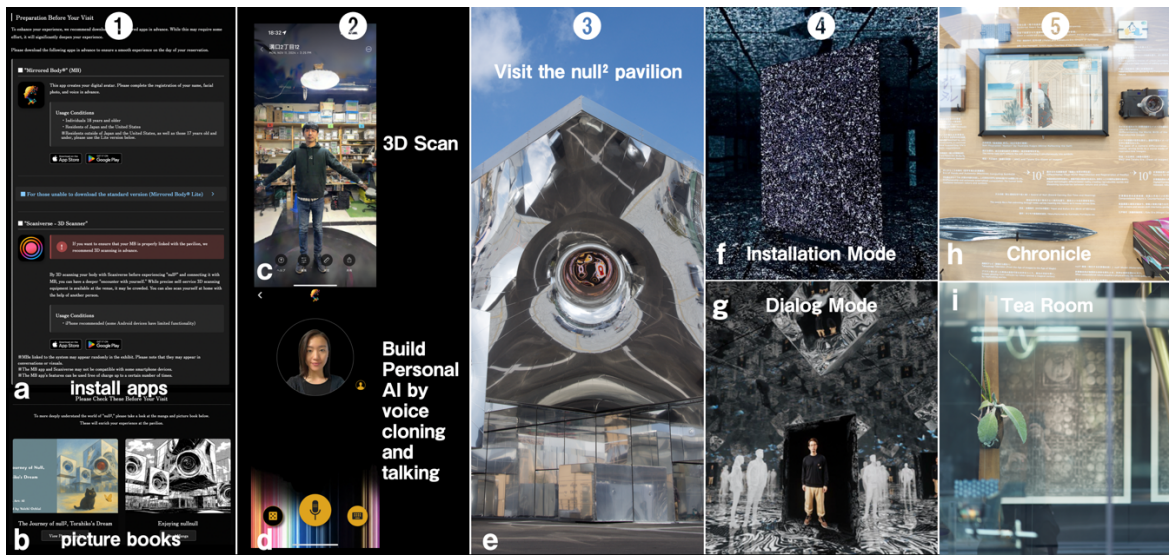

Figure 3. Visitor journey through the null² experience： Step 1 — Preparation before your visit: a Install apps: download the companion applications Mirrored Body and Scanniverse 3-D Scanner for avatar registration. b Picture books: read the digital picture-book series that introduces the narrative universe of null². Step 2 — Create your digital self: c 3-D Scan: use Scanniverse to capture a full-body three-dimensional scan. d Build Personal AI: perform voice cloning in the app to generate a conversational personal AI that will animate your avatar. Step 3 — Visit the null² pavilion: e Exterior: enter through the breathing mirror-membrane façade of the pavilion. Step 4 — Immersive modes inside: f Installation Mode: an all-enveloping generative environment of algorithmic particles and spatial sound. g Dialog Mode: the Mirrored-Body theatre, where visitors engage in real-time conversation with their own AI avatar. Step 5 — Post-experience zones: h Chronicle: a gallery that visualises the cumulative behavioural archive and artefacts produced by all visitors. i Tea Room: a contemplative space that fuses a traditional tea-ceremony ambience with digital-nature aesthetics.

The second experience, named Dialogue Mode, emphasizes interactions between visitors and their pre-generated digital twins. Visitors engage in real-time dialogues with their personalized AI and participate in narrative-driven scenarios embedded within the pavilion. Dialogue Mode further includes interactive experiences involving artificial lifeforms, transcending mere reflective imagery or simple dialogue. These interactions facilitate novel perceptual relationships between self and environment, extending existing research on self-representation and digital embodiment in social virtual reality [19] by manifesting these concepts in a physically tangible, spatially engaging form. In this mode, visitors first view the mirrored interior chamber from within, followed by an external viewing, thereby enhancing the experiential depth through this duality.

After concluding their internal pavilion experiences, visitors exit into spaces designated for reflective contemplation and historical contextualization, specifically a modernized interpretation of a traditional tea room and a chronological exhibit. The tea room environment, inspired by traditional Japanese cultural practices such as Zen Buddhism, offers visitors a serene space to reflect upon their digital and physical experiences. This area symbolizes the fusion of traditional Japanese spirituality with contemporary digital installation art, illustrating how Eastern philosophical perspectives and traditions can be reinterpreted within contemporary digital contexts. Simultaneously, the historical timeline exhibit articulates humanity's technological, spiritual, and cultural evolution, culminating in AI-oriented visions of future society, thus visually and contextually connecting null² with broader historical and cultural narratives.

In this manner, null² achieves a comprehensive visitor experience from pre-arrival through post-visit phases, transcending conventional architectural exhibitions by fostering deep self-dialogue, bodily interaction, and reflective engagement with philosophical and historical contexts.

## 4 CONSTRUCTION PROCESS

The basic plan provided by the Japan Association for the 2025 World Exposition explicitly defines Signature Pavilions as "organically transforming mirror-sculptures that actively reshape their surrounding landscape" and highlights the importance of creating dynamic, interactive relationships with visitors [25]. In alignment with this vision, null² embodies an architectural concept based on kinetic mirror membranes that continuously reflect and respond in real-time to the environment and visitor movements. This approach transcends traditional static architectural frameworks, enabling an organic, fluid interaction that makes the pavilion itself seem alive and responsive.

At the heart of realizing this concept is the Dynamic Mirror Membrane technology, a composite material combining a resin substrate with metal deposition [20]. Precise dynamic control of this membrane is achieved through a combination of advanced actuation mechanisms, including robotic arms for detailed spatial control, subwoofer-type vibrational devices providing low-frequency vibrations, localized tapping actuators delivering precise vibrational impulses, and linear actuators facilitating linear membrane expansion and contraction [21][22]. This intricate combination of actuation technologies enables subtle and responsive deformations of the membrane, generating interactive visual effects synchronized with visitor movements and environmental changes.

Structurally, the pavilion utilizes a voxel-based modular design. Voxel structures, wherein architectural forms are divided into cubic units, ensure high flexibility for planning modifications while visually expressing digital structural units [23].

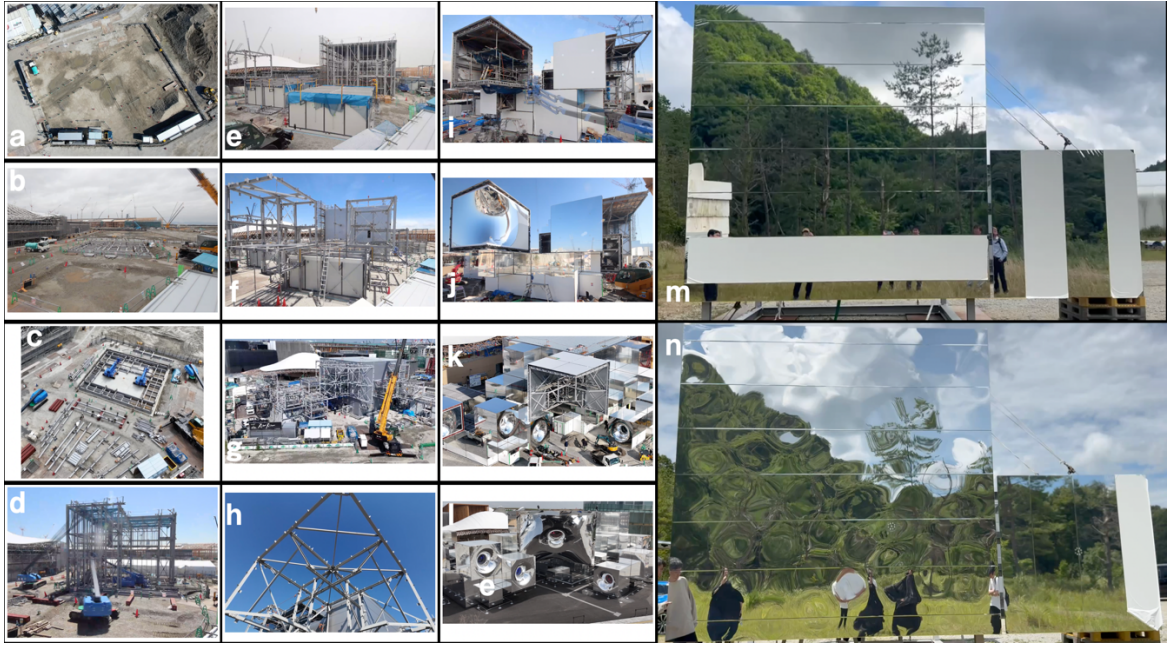

Figure 4. Construction process of the null² pavilion. (A) Cleared and segmented site. (B) Theater foundation construction begins. (C) Theater foundation completed, starting building construction. (D) Theater steel framework assembled. (E) Completion of theater framework, installation of prefab units for management and security buildings. (F) Facade structural assembly initiated. (G) Facade structure nearing completion. (H) Main facade structure completed. (I) Mirror membrane installation begins. (J) Quality check of installed mirror membrane. (K) Mirror membrane installation nearly complete. (L) Pavilion fully completed. (M) Outdoor weather durability test of mirror membrane. (N) Vibration testing of mirror membrane.

Each voxel unit independently controls the deformation of the mirror membrane, enabling dynamic visual expression responsive to visitor interactions and environmental variations. Such flexibility and interactivity position null² at the forefront of current advancements in responsive architectural environments [24].

To enhance immersive experiences within the internal space, an infinite reflection structure has been integrated. The pavilion's floor consists of specialized half-mirror sheets with approximately 50% reflectivity combined with tempered glass layers. LED lighting embedded beneath this floor creates vertical infinite reflections. Unlike conventional infinite reflection installations such as Yayoi Kusama's "Infinity Mirror Room" [17], the embedded LED lighting within the mirror surface generates complex and multi-layered visual effects.

The construction process (shown in figure 4 (a)-(n)) involved assembling a voxel-based steel frame during initial stages, upon which the dynamic mirror membrane was precisely installed. Subsequently, various actuation mechanisms were integrated to optimize membrane behavior and visual effects. The internal infinite reflection space was then completed by installing composite layers of half-mirror sheets and tempered glass, integrated with an underlying LED lighting system. This multidisciplinary construction approach required close collaboration among architectural engineers, media artists, and technical specialists, exemplifying a highly integrated implementation process (shown in figure 5 (a)-(h)).

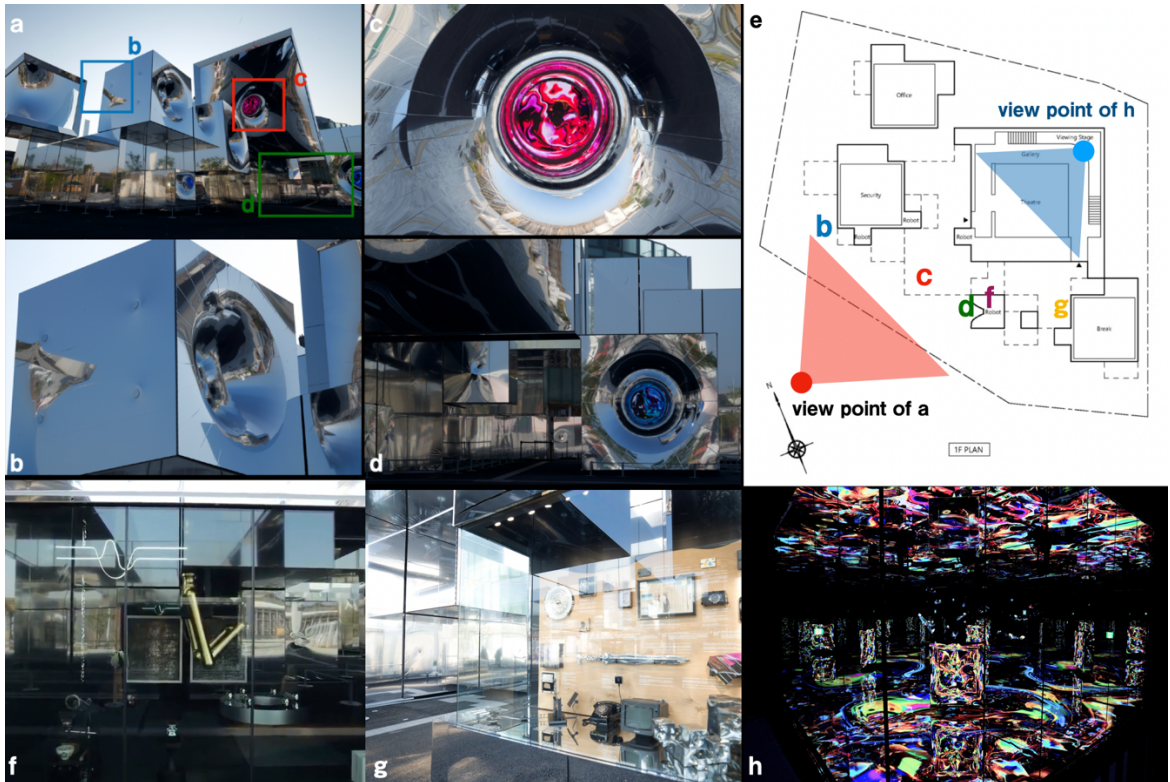

Figure 5. Exterior facade, dynamic structural details, and architectural layout of the null² pavilion. (A) Pavilion exterior facade. (B) Mirror membrane facade actively deformed by robotic arms. (C) LEDs installed within the horn-shaped curvature of the facade. (D) Membrane deformation driven by robotic arms and acoustic vibrations. (E) Architectural plan of the pavilion. (F) Interior view of the voxel-structured tea room located behind the membrane section shown in (D). (G) Cutaway view illustrating the Chronicle historical timeline exhibition. (H) Viewpoint location from the outer corridor surrounding the theater

## 5 INTERIOR AND INTERACTION

The interior space of null² is designed to provide visitors with an immersive experience integrating visual, auditory, and bodily sensations through advanced interactivity. At the center of the space is a movable LED monolith, approximately two meters in height, surrounded at its upper section by an interactive device termed the "moving sacred body," composed of robotic arms. At the tip of these robotic arms are mirror cubes constructed from the same materials as the external facade, equipped with actuators and subwoofers enabling precise movements. These cubes dynamically respond to both programmed scenarios and visitor interactions, creating compelling visual expressions [27]. Additionally, the floor incorporates LiDAR sensors to detect the real-time positions of visitors, using shader code to alter visual representations of artificial lifeforms dynamically, thereby enhancing the organic and interactive sensation throughout the space [28]. Collectively, these components create a highly immersive experience, wherein visitors perceive the entire environment as fluidly interacting with their presence.

The visitor experience within null² is structured around two distinct modes: "Installation Mode" and "Dialogue Mode." In "Installation Mode," visitors engage with the environment by observing the mirror cube installation from outside shown in figure 5 (e): view point (h). In this mode, the walls, floor, and ceiling incorporate mirrored surfaces embedded with full

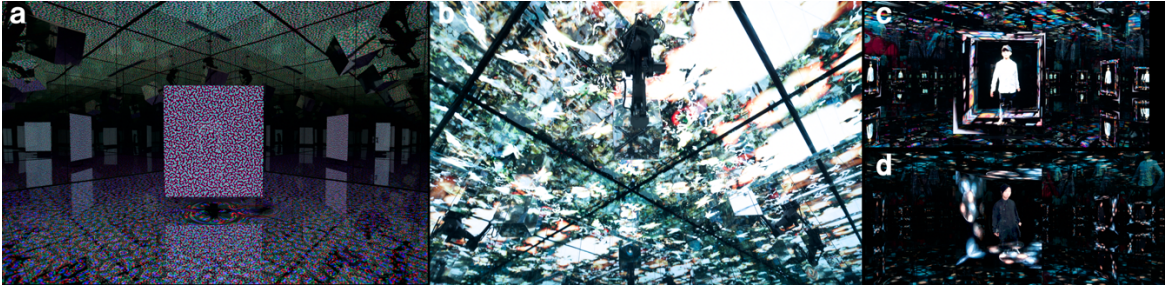

Figure 6. Exhibition scenes inside the null² pavilion theater. (A) Artificial life forms (Lenia) and monolith. (B) Robotic arm-controlled mirror cubes synchronized with ceiling LEDs. (C) Example of a digital human displayed on the monolith, with generative visuals above and below, customized according to individual interests and preferences. (D) Example of a Mirrored Body displayed on the monolith, also featuring generative visuals customized to individual interests and preferences.

LED arrays, producing a high-contrast infinite reflection that distinguishes itself significantly from conventional mirror installations [29]. The interactive movements of the monolith and robotic sacred body, combined with synchronized visuals and illumination effects, render the entire space as a living, responsive entity.

Conversely, in "Dialogue Mode," visitors enter the interior of the mirror cube itself shown in figure 6 (a)-(d). Within this space, visitors experience predetermined narrative scenarios that dynamically integrate real-time digital avatars of themselves. These digital representations are generated based on the visitors' physical appearance, voice, and personality traits, enabling real-time interactive dialogues between visitors and their digital counterparts [26]. Notably, three visitors are randomly selected as primary subjects for personalized interaction, with their digital avatars displayed live on the monolith, reflecting their own voices and personal information in real time.

This dialogue-driven interaction is processed through real-time generative artificial intelligence powered by large language models (LLMs), extracting conversational content, preferences, and personal traits from interactions to produce personalized audiovisual content instantaneously. Similar interactive techniques leveraging LLM-driven narrative generation and dynamic audiovisual feedback have been demonstrated in recent media art installations, notably ReCollection, where visitor conversations directly inform real-time generated imagery and soundscapes [30]. As conversations progress, visuals reflecting the inner attributes of visitors are dynamically generated within the space, creating uniquely personalized narratives for each participant. Through this interactive process, visitors directly experience the symbolic decomposition and reconstitution of their identities and inner selves in real-time, both visually and aurally.By seamlessly merging sophisticated interactive spatial designs with narrative-driven experiential scenarios, null² surpasses traditional static exhibitions and interactions, offering visitors a profound immersive experience that prompts them to reconsider their own existence and bodily perceptions. Positioned as an exploration of new ontological understandings, null² enables visitors to engage deeply with organic artificial lifeforms and digital-human representations of themselves within an endlessly expanding digital environment.

# 6 MIRRORED BODY APP

The Mirrored Body is a core mobile application supporting visitor experiences at the null² pavilion shown in figure 7 (a)-(e). It enables daily interactions with visitors' digital twins, aiming to philosophically and technologically redefine self-awareness and bodily perception. The underlying philosophical framework of this app derives from Buddhist teachings on "emptiness" (śūnyatā), particularly the Heart Sutra's concept of "form is emptiness, emptiness is form" [9]. The concept of emptiness negates any fixed, substantial self, suggesting the dismantling of rigid self-images in favor of perceiving oneself as a dynamic and fluid entity. Mirrored Body translates this philosophy into digital technology by precisely scanning visitors' bodies, symbolizing them digitally as avatars, and thus offering new pathways for self-understanding.

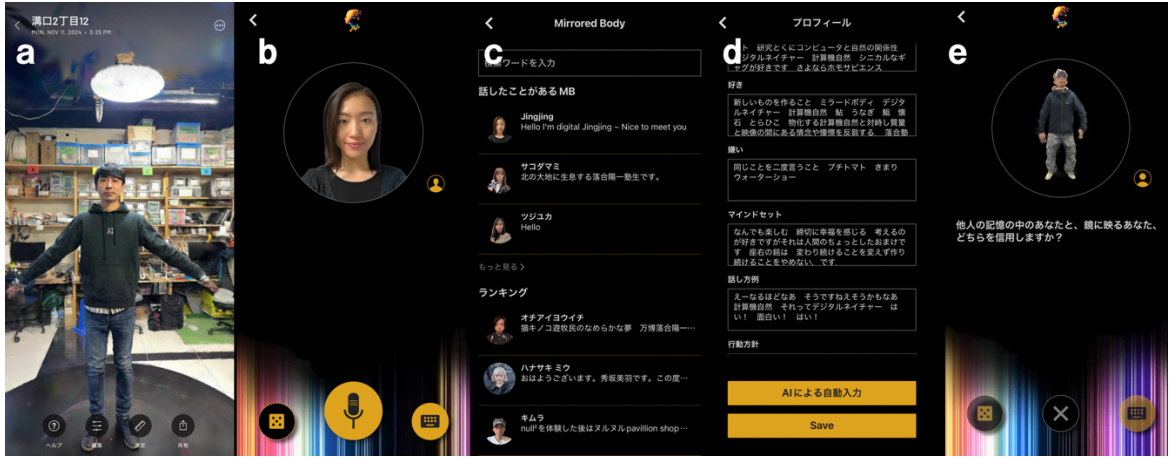

Figure 7. Mirrored Body generation and conversational application interface and functions. (A) Gaussian Splatting of user's 3D scan captured via Scanniverse. (B) Mirrored Body app user interface (includes speak button, random talk button, keyboard input options). (C) Interface allowing users to select and converse with other visitors' Mirrored Bodies. (D) User profile collected from interactions with the app, utilized for conversation generation. (E) Demonstrating real-time facial deformation and lip-syncing during speech.

The specific user experience with Mirrored Body begins with visitors installing dedicated mobile applications, Mirrored Body and Scaniverse, on their smartphones. Visitors then perform detailed 3D scans of themselves using the Gaussian Splatting technology integrated into Scaniverse. Gaussian Splatting is a state-of-the-art technique that enables real-time, high-resolution rendering of dense 3D point clouds, known particularly for its effectiveness in digitizing human figures [31][32]. Digital avatars generated from this scanning process faithfully clone the user's physical attributes, voice, and facial expressions. Moreover, users' personality traits are precisely modeled through AI dialogue algorithms, enabling continuous daily interactions. This sustained self-interaction constitutes a significant innovation of Mirrored Body [14][33]. Recently, many services using digital avatar generation have emerged. However, most of these existing systems only offer simple customization of appearances or temporary interactions [19]. In contrast, Mirrored Body allows for continuous daily interactions with digital twins authentically cloned from users' actual appearances and voices. Furthermore, user data gathered through these interactions accumulates persistently, enabling real-time feedback reflecting users' evolving inner traits. This process fosters deepened self-awareness, permitting visitors to externalize their bodily symbols digitally and deepen their self-understanding. When visitors physically enter the pavilion, the digital twin data stored securely on their smartphones synchronizes in real-time with the pavilion's interactive systems, projecting their digital selves into the exhibit's interactive environment. To ensure secure daily management of digital avatar data, Mirrored Body employs blockchain technology, providing secure, decentralized data management. This design prevents unauthorized data modification or misuse, aligning with UNESCO's recommendations on ethical AI principles, particularly those emphasizing personal data protection and transparency [34][35].

## 7 SCENARIO AND CONTEXT

The null² pavilion provides a multilayered, interconnected experience through its historical timeline exhibited externally, interactive narrative scenarios unfolding internally, and a tea room space that reinterprets traditional culture digitally, collectively offering visitors profound philosophical and cultural contexts.

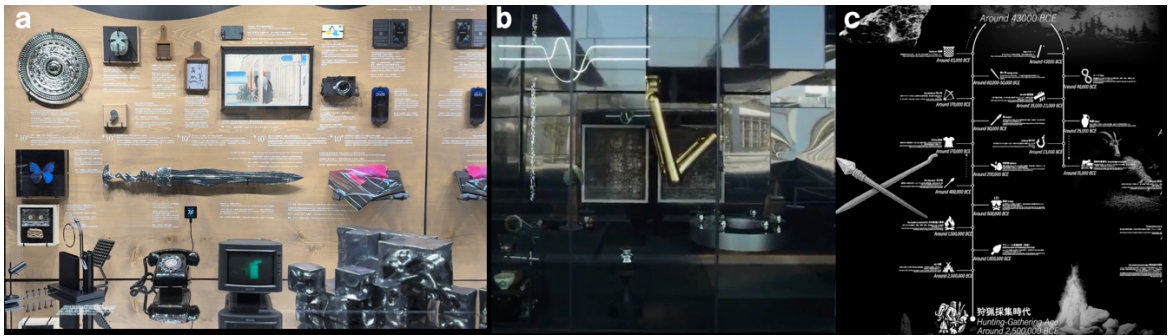

Figure 8. Historical and cultural context exhibits within the null² pavilion. (A) Exterior exhibition of historical timeline and physical artifacts illustrating exponentially accelerating technological evolution, from the era of $10^{-5}$ to $10^0$. (B) Tea-room exhibit employing motifs from Japanese culture. The golden robotic arm pays homage to Toyotomi Hideyoshi's golden tea room; the mandala displayed in the background references similarities with the pavilion's mirrored structure. (C) Excerpt from the historical timeline of human technological progress used within the pavilion's theater content, presented after the artificial-life experience.

The external space (shown in figure 8 (a)) features a chronological exhibition of human history depicted on a logarithmic scale. This display visually and physically immerses visitors in the progression from humanity's origins through key historical milestones—such as the discovery of fire, the birth of religions, advancements in tools and sciences—up to contemporary society increasingly permeated by artificial intelligence and digital technologies [36]. Employing a logarithmic scale allows visitors to intuitively grasp the accelerating rate of historical change towards the present [37].

Linked seamlessly with the external exhibition, the pavilion's internal Dialogue Mode offers visitors an interactive narrative experience centered around their digital avatars (Mirrored Bodies). This narrative poetically symbolizes human cultural and technological evolution, depicting milestones such as the mastery of fire, the emergence of religion, tool-making, scientific advancements, and eventually projecting into futures characterized by digital nature (shown in figure 8 (c)). Designed to interweave the visitors' personal interiority with a broad context of human history, visitors engage in real-time reinterpretations of their symbolic selves, confronting deep philosophical inquiries [38].

Additionally, within the pavilion, the tea room (shown in figure 8 (b)) space represents a contemporary digital interpretation of the traditional Japanese tea ceremony. Drawing inspiration from Toyotomi Hideyoshi's historically significant Golden Tea Room, this space employs robotic arms and advanced 3D printing technology in its modern reconstruction [39][40]. Specific elements include the "Pla-Raku," a contemporary rendition of traditional Raku tea bowls produced through 3D printing technology [41], and artworks such as the Daigoji-Mandara created with high-definition platinum printing [42]. These artistic innovations present visitors with a fresh fusion of traditional culture and digital technologies, effectively redefining the spiritual essence of the tea ceremony within the framework of digital art and exemplifying the integration of Eastern philosophy with contemporary technological practices [43].

## 8 DISCUSSION, CONCLUSION, AND FUTURE WORK

This paper provides a rapid, comprehensive case study of the artistic, architectural, and technological fusion embodied in the null² pavilion. Due to its exploratory nature, the detailed technical specifics and design processes have been limited; deeper technical discussions and thorough evaluations will be addressed in subsequent publications. Instead, this paper emphasizes the overarching philosophical and artistic concepts, considering it essential to articulate clearly the ideological and experiential significance arising from the fusion of art, architecture, and technology that null² represents.

A notable feature of null² lies in its architectural conception as a dynamically transformable sculptural object, intricately connected with visitors' experiential interactions both externally and internally. In particular, interactions between artificial life forms inhabiting the internal pavilion space and visitors' digital avatars present narratives involving the symbolic "coding" and "decoding" of the self. This approach surpasses conventional visual and spatial experiences, provoking deeper philosophical and narrative introspection. Compared to traditional mirrored works and digital installations such as Anish Kapoor's Sky Mirror [44] or Refik Anadol's Machine Hallucinations [45], null² provides a uniquely personal, symbolic, and interactive experience. Its integration of real-time AI interactions, delving profoundly into visitors' philosophical and psychological interiors, highlights a critical novelty and reflective dimension lacking in previous works.

Nevertheless, clear limitations exist within this case study format. Technical details regarding the precise design process, material specifications, and control methodologies essential to dynamic architectural design remain inadequately covered. Moreover, quantitative evaluations of user experiences and empirical visitor behavior data, particularly concerning the Mirrored Body application, fall outside the scope of this study. Examining variations in visitor experiences based on cultural backgrounds and psychological characteristics is particularly crucial for understanding the multifaceted, philosophical dimensions that the null² art project seeks to foster.

Future research should encompass detailed technical follow-ups regarding the dynamic mirror membrane technology, including specific design methods and control systems. Additionally, empirical research evaluating visitors' quantitative and qualitative experiences will be essential. Developing and refining criteria and methodologies for assessing art-technology hybrid projects that stimulate philosophical introspection is another recognized need. This initial report deliberately refrains from extensive technical exposition, instead laying a foundational overview of the project's philosophical and conceptual dimensions, thereby providing a platform for subsequent detailed academic exploration.

**REFERENCES**

[1] Greenhalgh, P. 1988. Ephemeral Vistas: The Expositions Universelles, Great Exhibitions and World's Fairs, 1851-1939. Manchester University Press, Manchester, UK.

[2] Winther-Tamaki, B. 2011. To Put On a Big Face: The Globalist Stance of Okamoto Tarō's Tower of the Sun for the Japan World Exposition, 1970. Review of Japanese Culture and Society 23 (Dec.), 81-101.

[3] Japan Association for the 2025 World Exposition (公益社団法人 2025 年日本国際博覧会協会). 2025. Outline of Thematic Project Pavilion at Expo 2025 Osaka, Kansai, Japan (in Japanese). Retrieved June 24 2025 from https://www.expo2025.or.jp/en/project/

[4] Ochiai, Y. 2018. デジタルネイチャー——生態系を為す汎神化した計算機による侘と寂 (Digital Nature: Wabi-Sabi as Computation-Integrated Panentheism Forming an Ecosystem). PLANETS Publishing, Tokyo, Japan (in Japanese).

[5] Ochiai, Y. 2024. SINIC Theory and Digital Nature. IEICE Transactions (The IEICE Journal) 107, 3 (Mar.), 226-232. DOI: 10.14923/ieice.107.3.226

[6] Honisch, P. 2024. Making history, making place—contextualising the built heritage of world expos 2010 and 2015. Built Heritage 8, Article 35 (Aug. 2024). DOI: 10.1186/s43238-024-00143-2.

[7] Isozaki, A. 1979. MA: Space–Time in Japan. Exhibition catalogue, Cooper-Hewitt Museum, Smithsonian Institution, New York, NY.

[8] Engel, H. 1985. The Japanese House: A Tradition for Contemporary Architecture. Charles E. Tuttle Co., Rutland, VT / Tokyo, Japan. (Orig. ed. 1964.)

[9] Laycock, S. W. 1994. Mind as Mirror and the Mirroring of Mind: Buddhist Reflections on Western Phenomenology. State University of New York Press, Albany, NY.

[10] Kuo, M. 2018. "'To Avoid the Waste of a Cultural Revolution': Experiments in Art and Technology (E.A.T.), 1966–1974." Ph.D. dissertation, Department of History of Art and Architecture, Harvard University. ProQuest Dissertations & Theses Global (No. 28223005).

[11] Expo 2020 Dubai. 2020. Switzerland Pavilion – "Reflections". Country-pavilion dossier. Retrieved June 24 2025 from https://www.expo2020dubai.com/en/understanding-expo/participants/country-pavilions/switzerland

[12] Hein, J. 2011. Mirror Wall. Interactive installation. In Gesamtkunstwerk: New Art from Germany (Saatchi Gallery, London, UK, 18 Nov. 2011 – 15 Apr. 2012). Exhibition catalogue, Saatchi Gallery, London.

[13] Rozin, D. 1999. Wooden Mirror. Interactive sculpture (830 wooden tiles, servo motors, video-camera vision system). Artist documentation. Retrieved June 24 2025 from https://www.smoothware.com/danny/woodenmirror.html

[14] Wang, R., Chen, C.-F., Peng, H., Liu, X., Liu, O., and Li, X. 2019. Digital Twin: Acquiring High-Fidelity 3D Avatar from a Single Image. arXiv preprint arXiv:1912.03455. DOI: 10.48550/arXiv.1912.03455.

[15] Li, J., Wider, W., Ochiai, Y., and Fauzi, M. A. 2023. A bibliometric analysis of immersive technology in museum exhibitions: Exploring user experience. Frontiers in Virtual Reality 4, Article 1240562. DOI: 10.3389/frvir.2023.1240562.

[16] Bullivant, L. 2006. Responsive Environments: Architecture, Art and Design. Victoria & Albert Museum, London, UK. ISBN 978-1-85177-481-4.

[17] Yoshitake, M. (ed.). 2017. Yayoi Kusama: Infinity Mirrors. Exhibition catalogue. DelMonico Books/Prestel in association with the Hirshhorn Museum and Sculpture Garden, Washington, DC / Munich, Germany. ISBN 978-3791355948.

[18] teamLab. 2018. teamLab Borderless – Exhibition Documentation. MORI Building Digital Art Museum: EPSON teamLab Borderless, Tokyo. Retrieved June 24 2025 from https://borderless.teamlab.art/

[19] Freeman, G. and Maloney, D. 2020. Body, Avatar, and Me: The Presentation and Perception of Self in Social Virtual Reality. Proceedings of the ACM on Human-Computer Interaction 4, CSCW3, Article 239 (Dec. 2020), 27 pages. DOI: 10.1145/3432938.

[20] Gramazio, F. and Kohler, M. 2008. Digital Materiality in Architecture. Lars Müller Publishers, Baden, Switzerland. ISBN 978-3-03778-122-7.

[21] Hosseini, S. M., Mohammadi, M., and Guerra-Santin, O. 2019. Interactive kinetic façade: Improving visual comfort based on dynamic daylight and occupant's positions by 2D and 3D shape changes. Building and Environment 165, Article 106396. DOI: 10.1016/j.buildenv.2019.106396.

[22] Elkhayat, Y. O. 2014. Interactive movement in kinetic architecture. Journal of Engineering Sciences, Assiut University 42, 3 (May 2014), 816-845. DOI: 10.21608/jesaun.2014.115027.

[23] Terzidis, K. 2006. Algorithmic Architecture. Architectural Press (Elsevier), Oxford, UK. ISBN 978-0-7506-6725-8.

[24] Bullivant, L. 2006. Responsive Environments: Architecture, Art and Design. Victoria & Albert Museum (V&A Contemporary Series), London, UK. ISBN 978-1-85177-481-4.

[25] Japan Association for the 2025 World Exposition (公益社団法人 2025 年日本国際博覧会協会). 2022. 2025 年日本国際博覧会 基本計画 (Master Plan for Expo 2025 Osaka‑Kansai) (in Japanese). Osaka, Japan. Retrieved June 24 2025 from https://www.expo2025.or.jp/overview/masterplan/

[26] Wang, R., Chen, C.-F., Peng, H., Liu, X., Liu, O., and Li, X. 2019. Digital twin: Acquiring high-fidelity 3-D avatar from a single image. arXiv preprint arXiv:1912.03455. DOI: 10.48550/arXiv.1912.03455.

[27] Hosseini, S. M., Mohammadi, M., and Guerra-Santin, O. 2019. Interactive kinetic façade: Improving visual comfort based on dynamic daylight and occupant's positions by 2D and 3D shape changes. Building and Environment 165 (Sept.), Article 106396. DOI: 10.1016/j.buildenv.2019.106396.

[28] Wojciechowski, R., Walczak, K., White, M., and Cellary, W. 2004. Building virtual and augmented reality museum exhibitions. In Proceedings of the 9th International Conference on 3D Web Technology (Web3D '04, Monterey, CA, USA, 5–8 Apr. 2004). ACM, New York, NY, 135–144. DOI: 10.1145/985040.985060.

[29] Sommerer, C. and Mignonneau, L. 1999. Art as a living system: Interactive computer artworks. Leonardo 32, 3 (June), 165–173. DOI: 10.1162/002409499553190.

[30] Zhang, W., Cheng, L., and Luo, J. 2024. ReCollection: Creating synthetic memories with AI in an interactive art installation. Proceedings of the ACM on Computer Graphics and Interactive Techniques 7, 4, Article 51 (Jul.). DOI: 10.1145/3664207.

[31] Kerbl, B., Kopanas, G., Leimkühler, T., and Drettakis, G. 2023. 3D Gaussian Splatting for Real-Time Radiance Field Rendering. ACM Transactions on Graphics 42, 4, Article 146 (July). DOI: 10.1145/3592433.

[32] Niantic Inc. 2023. Scaniverse: LiDAR 3-D Scanning App with Gaussian Splatting. Mobile application website. Retrieved June 24 2025 from https://scaniverse.com

[33] Sustainable Pavilion 2025 Inc. 2025. Mirrored Body®: Interactive Digital Avatar Platform. Retrieved June 24 2025 from https://sp-2025.com/en/mirrored-body

[34] Zheng, Z., Xie, S., Dai, H., Chen, X., and Wang, H. 2018. Blockchain challenges and opportunities: A survey. International Journal of Web and Grid Services 14, 4, 352–375. DOI: 10.1504/IJWGS.2018.095647

[35] UNESCO. 2021. Recommendation on the Ethics of Artificial Intelligence. Paris: United Nations Educational, Scientific and Cultural Organization. Retrieved June 24 2025 from https://unesdoc.unesco.org/ark:/48223/pf0000380455

[36] Christian, D. 2004. Maps of Time: An Introduction to Big History. University of California Press, Berkeley, CA. ISBN 978-0-520-24476-4.

[37] Aigner, W., Miksch, S., Schumann, H., and Tominski, C. 2011. Visualization of Time-Oriented Data. Springer, London, UK. ISBN 978-0-85729-078-6.

[38] Manovich, L. 2002. The Language of New Media. MIT Press, Cambridge, MA. Paperback ed., ISBN 978-0-262-63255-3.

[39] Isozaki, A. 2006. "The Golden Tea Room." In Chado: The Way of Tea—A Japanese Tea Master's Almanac, S. Sanmi (ed.), Tuttle Publishing, North Clarendon, VT, 88–93.

[40] Kusahara, M. 2000. Presence, absence, and knowledge in telerobotic art. In The Robot in the Garden: Telerobotics and Telepistemology in the Age of the Internet, K. Goldberg (ed.), MIT Press, Cambridge, MA, 198-212.

[41] Ochiai, Y. 2021. 可塑庵 (Pla-An): Plastic Raku Tea Hut. Installation presented at "Gundam R (Recycle) Operation FINAL 2021," Shinjuku Sumitomo Building Triangle Square, Tokyo, Japan, 20 – 21 Nov. 2021. Exhibition documentation. Retrieved June 24 2025 from https://yoichiochiai.com/exhibition/gundam-recycling-operation-2021/.

[42] Ochiai, Y. 2021. Imaging Mandala Study V (Daigoji-Mandara). Platinum / palladium print on aluminium composite panel. Shown in the solo exhibition "The Transformation of Material Things into a Living Forest – Digital Nature –," Daigoji Temple, Kyoto, Japan, 21 Nov.–5 Dec. 2021. Exhibition documentation retrieved June 24 2025 from https://yoichiochiai.com/art/imaging-mandala-study/

[43] Yanagi, S. 2000. 茶と美 (Cha to Bi－The Beauty of Tea) (in Japanese). 講談社学術文庫, Tokyo, Japan. ISBN 978-4-06-159453-1.

[44] Kapoor, A. 2006. Sky Mirror. Stainless-steel outdoor sculpture (first realised 2001; Rockefeller Center installation Sept.–Oct. 2006). Exhibition documentation, Public Art Fund, New York, NY. Retrieved June 24 2025 from https://www.publicartfund.org/exhibitions/view/sky-mirror/

[45] Anadol, R. 2019. Machine Hallucinations: NYC. Immersive AI-generated media installation, ARTECHOUSE, New York, NY (6 Sep – 1 Dec 2019). Exhibition documentation. Retrieved June 24 2025 from https://refikanadol.com/works/machine-hallucination-nyc/